\documentclass[useAMS,usenatbib]{mn2e}
\usepackage{graphicx}
\usepackage{amsmath}
\title[Low mass variable stars in the globular cluster NGC\,6397]{Low mass variable stars in the globular cluster NGC\,6397}

\author[E. Martinazzi, S. O. Kepler, J. E. S.~Costa]{E.~Martinazzi$^{1,2}$\thanks{E-mail: elizandra.martinazzi@ufrgs.br}, S.~O.~Kepler$^{1}$, J.~E.~S.~Costa$^{1}$\\
$^{1}$Instituto de F\'{\i}sica, Universidade Federal do Rio Grande do Sul, 91501-900 Porto Alegre, RS, Brazil\\
$^{2}$Instituto Federal do Rio Grande do Sul, 95700-000, Bento Gon\c{c}alves, RS, Brazil \\}

\begin{document}

\date{Accepted  Received XXX XXX XXX}

\pagerange{\pageref{firstpage}--\pageref{lastpage}} \pubyear{2015}

\maketitle

\label{firstpage}

\begin{abstract}

We have conducted a photometric survey of the globular cluster NGC\,6397 in a search for variable stars.
We obtained $\sim$ 11\,h of time-resolved photometric images with one ne European Southern Observatory-Very Large Telescope using the FOcal Reducer and low dispersion Spectrograph imager distributed over two consecutive nights.  We analyzed 8\,391 light curves of stars brighter than  magnitude 23 with the 465 nm-filter, and we identified 412 variable stars, reaching $\sim$ 4.8 $\pm$ 0.2 per cent of variability with timescales between 0.004 and 2\,d, with amplitudes variation greater than $\pm$ 0.2 mag.

\end{abstract}


\section{Introduction}

Stellar variability depends on effective temperature, magnetic field and also on the opacity, which is metallicity dependent.
Globular clusters offers a way to study this dependence. 
As globular cluster are some of the oldest objects in the Universe, they are laboratories for the study of the early stages of Galaxy formation \citep[e.g.][]{1998gaas.book.....B,2002ARA&A..40..487F}.
Each cluster is made up by a essentially simple population of stars, i.e. practically all stars were born from the same molecular cloud \citep[e.g.][]{Rosenberg2000} even though there is evidence for multiple populations in a few clusters.  The comparison between the ratio of variable M stars in function of their masses for globular clusters with different metallicities and the ratio for the galactic disk can tell us whether metallicity plays an important role at low temperatures.

The globular cluster NGC\,6397 is particularly interesting in this context. NGC\,6397 is a globular relatively easy to study due to its proximity. It is considered, together with NGC\,6101 (M4), one  of the two nearest clusters from the Sun, located at a distance $R_{SS}= 2.2^{+0.5}_{-0.7}$ kpc \citep{Heyl2012} at $\alpha = 17h\,40m\,42.09s$ and $\delta = -53^{\circ}40'27.6"$ (J2000), with galactocentric coordinates $\ell = 338.17^{\circ}$ and $b = -11.96^{\circ}$.  
NGC\,6397 is classified as a core-collapsed cluster, predicted by theory decades ago and confirmed by observations \citep{Trager1995,2014MNRAS.442.3105M}. 
The distance moduli for NGC\,6397 is $12.02 \pm 0.06$ \citep{2007ApJ...671..380H} with reddening  E(B-V) = $0.18\pm0.02$ [\cite{Harris1996} 2010 edition]. 
Another remarkable feature of NGC\,6397 is its low metallicity. In fact, it is one of the lowest metallicity globular clusters known, with $[Fe/H] = -1.99\pm 0.02$ \citep{Carretta2009}.

\cite{1996A&AS..120...83K} reported five new variable stars, including SX Phe stars and RR Lyrae variables. \cite{Kaluzny1997} reported that few variable stars are known in the globular cluster NGC\,6397. They obtained light curves for seven additional discoveries, identified as close binaries, SX Phe stars and RR Lyrae variable. \cite{2003AJ....125.2534K}, studied the central region of NGC\,6397 and detected nine new variables stars, including one eclipsing binary, new SX Phe stars and low amplitude variables. \cite{2006MNRAS.365..548K} presented 12 new objects, of which six are periodic light curves and eclipsing binaries of unknown period again in the central region of the cluster. 

\cite{2010ApJ...722...20C} studied X-ray sources identified by the Chandra telescope near the center of the cluster. In addition to studying nine cataclysmic variables (CVs) previously detected, they identified six new weak candidates for CV. \cite{2012A&A...541A.144N} conducted a search for variables and planetary transits in NGC\,6397, exploring the images of the HST (Hubble Space Telescope). They analyzed 5\,078 light curves, including a selection of 2\,215 cluster-member M dwarfs selected on the proper motions and reported 12 new variable stars.
The great majority of the variables previously known are in the central region of the cluster. 

The main objective of this work is to determine the variable star ratio as a function of mass for NGC\,6397, studying if the variability is related to the metallicity.

\section{Observational data and Data Reduction}

The images used in this work were obtained from the ESO program ID 083.D-0653(A) P.I. Barbara Castanheira. They were taken in 2009 July 27-28, with 8.4-m ESO-VLT-UT1 telescope (\textit{Very Large Telescope}), using the FORS2 (\textit{FOcal Reducer and low dispersion Spectrograph}) imager composed by two CCDs with a total of  $2\,048\times2\,068$ pixels and $0.25"$/pixel scale, with an usable portion total of $1\,670\times1\,677$ pixels. 

The images are taken with the filter FILT\_465\_250 (central wavelength $\lambda$=4\,650~\AA~ and $\Delta\lambda$=250~\AA~). A total of 305 images were obtained: 205 of 60\,s on the first night, 28 of 80\,s on the first part of the second night and 72 images of 170\,s on the second part of the last night, starting at Julian date JD = 2\,455\,038.507350. 

\begin{figure}
\centering
\includegraphics[width=84mm,clip=]{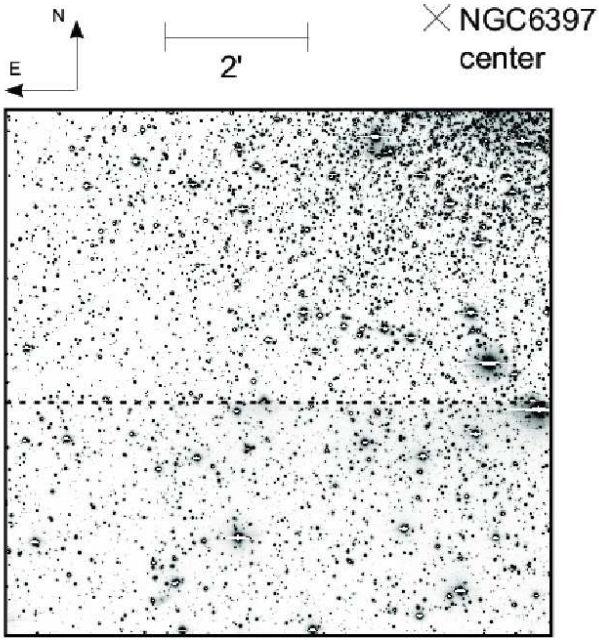}
\caption{ CCDs images of the NGC\,6397, CCD1 in the top and CCD2 in the bottom. The dotted line indicates the border between both CCDs. The top edge of the right CCD1 is distant $\sim1\arcmin$ from the cluster center ($0,72\,pc$).}
\label{ccds}
\end{figure}

We obtained 5.1\,h of photometry on the first night and 4.8\,h on the second night, with a separation of 19.95\,h. The CCD reading time was of the order of 30\,s.  CCD1 covers a $6,8'\times3,9'$ field and CCD2 $6,8'\times2,9'$ field, amounting to $6,8'\times6,8'$ the whole field, without any gap between the CCDs. The images were centred at $\alpha = 17h\,41m\,01.78s$ and $\delta = -53^{\circ}44'47.2"$ (J2000). The cluster centre, excluded from the images, is around 1.1\,arcmin to the upper right corner of the CCD1 (Fig.~\ref{ccds}).

Image data reduction were carried out with standard \textsc{IRAF} routines, with tasks \textit{daofind, phot} and \textit{allstar}, applying Point Spread Function (PSF) photometric procedures. 
The astrometric calibration was done using the coordinates of the white dwarfs  studied by \cite{2000A&A...354L..75M}.
The images were centered on R.A. =  $17h\,41min\,01,65\,s$ e DEC. = $ 53^{0}44'46,05"$ (J2000)  to cover the region where the white dwarf candidates for DAVs.
The correction for atmospheric extinction for a given filter was applied from the equation  $m_{0} = ZP + m - K X $,
where $ m_{0} $ is the corrected apparent magnitude, $ m $ is the apparent instrumental magnitude, $ ZP $ is the zero point, $ X \simeq sec (z) $, the air mass, which varies throughout the night, depending on the zenithal distance $z$ of the object.

To create a reference magnitude data and subtract from the measured magnitudes for a given image,
we compared the instrumental magnitude with the apparent magnitude of twenty standard cluster stars and determine a $ZP$ for each image. This procedure is equivalent to assuming that $ZP$ is constant and $K$  undergoes fluctuations over time.
For the photometric calibration, we use the standard stars in the same field, measured by \cite{Stetson2000} and \cite{Stetson2005}. After we applied the photometric corrections we obtained the light curves of each star, where each point of the light curve is the apparent magnitude calculated photometrically by PSF fitting. 
The extinction coefficient for the filter FILT\_465\_250 was calculated from an interpolation of tabulated values. The mean value assumed for this filter was $ K = $ 0.189. The average seeing was 1.18" for the first night and 0.69" for the second night.
We included, in the calculations, the effect of the uncertainties in the instrumental magnitude, the zero point and the air mass.

\begin{figure*}
\begin{minipage}{115mm}
\centering
\includegraphics[width=115mm,clip=]{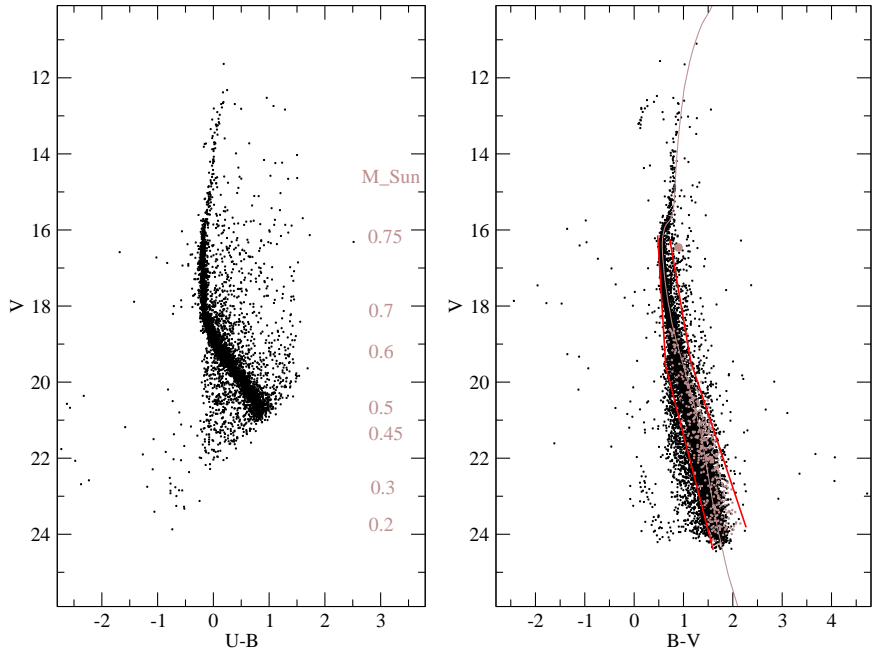}
\caption{ NGC\,6397 color-magnitude diagram for the data of this work. The values shown in the left panel represent the approximate position of the masses obtained by isochrones, line in the right panel, calculated by A. Dotter (2008) to the lower limit of the burning of H.  The lines at the right side CMD shown the colour-magnitude filter applied in order to exclude probable non-member objects as the field stars. }
\label{CMDubv}
\end{minipage}
\end{figure*}

We built the color-magnitude diagram (CMD) shown in Fig.\ref{CMDubv} from images obtained with B$\_HIGH$ and V$\_HIGH$ filters, with exposure times of 1, 10, 120, 300\,s each, and 1, 300 e 600\,s with  U$\_HIGH$ filter. The data reduction were performed using  \textsc{iraf} routines, with the tasks \textit{daofind, phot} and \textit{allstar}, applying standard point spread function  photometric procedures. The best-fitting PSF function among those available was the elliptical Moffat with coefficient $\beta = 2.5$.  Objects brighter than V$\_HIGH$ filter $\simeq $ 19 were saturated in the 300\,s images. Two lists of positions were created for each CCD using the images of 1\,s (V$\_HIGH$ filter)  for the brightnest stars and 300\,s  (V$\_HIGH$ filter) for the faint stars and both data sets were combined to avoid superposition.

In order to exclude probable non-member objects as the field stars, we use a colour-magnitude filter applied to the CMD built with the ESO-VLT data. This procedure eliminated $\sim$ 25 per cent of the stars originally present in the CMD of NGC\,6397. 
In \cite{2014MNRAS.442.3105M} the luminosity and density of Milky Way field stars in the direction of NGC\,6397 were estimated, obtaining a contamination by Galactic stars of 0.04 $stars/arcsec^{2}$.
The cut is shown in Fig.~\ref{CMDubv} at the right side CMD. 

\begin{figure}
\centering
\includegraphics[width=84mm,clip=]{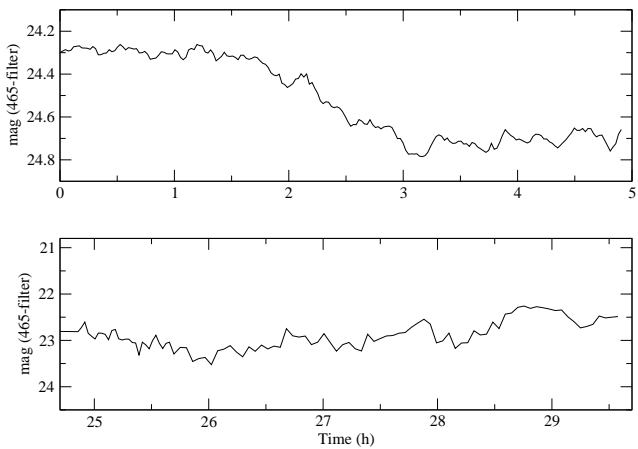}
\caption{ Brightness variation of the sky during both observation evenings in the 465nm-filter. In the first night (top) we used exposure time of 60\,s. In the second night (bottom), we used 80\,s in the fist part and 170\,s in second part of the observations. }
\label{sky3}
\end{figure}

We found 7\,617 objects in CCD1 and 3\,739 in CCD2, totalling 11\,356 objects in the whole field with magnitudes between 15 and 25. However, for  stars brighter than magnitude 23 in the 465nm-filter, we find a total 8\,391 stars. This range of magnitude was selected because fainter stars than magnitude 23 were contaminated by the large brightness variation of the sky (zodiacal light, even without moon) as shown in Fig.~\ref{sky3}. 
We only used light curves for stars detected in more than 100 images. We also eliminate the stars near the edge of images, over a distance of 15 pixels from the edges.

\section{Low mass variable stars}

The M dwarfs are the most numerous stars in the Galaxy \citep[e.g.][]{2014arXiv1409.0862R}. 
The understanding of the inner structure of the largest population of stars in our Galaxy could occur through observations of pulsating M dwarfs. \cite{2014MNRAS.438.2371R} presented a theoretical study  of the instability strip of M dwarf stars for models in the range 0.10 - 0.60\,M$_{\odot}$, with model for different metallicity  in the range -1.0 $ < [Fe/H] < $ 0.0, showing theoretical evidence that solar-like oscillations can be excited in these stars. They found  that the M dwarfs unstable modes have periods ranging from 0.014 - 0.46\,d depending on the mass and evolutionary stage.

Solar-like oscillations have small amplitudes, but variations were observed to be stronger than predicted in the planning by the Kepler satellite \citep{2005stam.book.....G}. 
Manifestations of the activity of lower main-sequence stars are numerous and diverse. Sporadic flares are observed in all layers of the stellar atmosphere, as well as cool spots involving the stellar surface and manifesting variability of large-scale structures in stellar atmospheres, chromospheres and coronae \citep{2005stam.book.....G}. All these effects could have longer timescale than we studied.

\cite{2011AJ....141..166H} investigated the optical broadband photometric variability of a sample of 27\,560 field K and M dwarfs selected by color and proper motion using light curves from the HATNet survey for transiting extrasolar planets. They searched the light curves for periodic variations and for large-amplitude, long-duration flare events totalling 2\,120 stars with potential variability using timebases between 45 days and 2.5 years. They also found that the rotation periods and amplitudes of K and early-to mid-M dwarfs are uncorrelated for periods less than 30 days, and that amplitude decreases with increasing rotation period greater than 30 days. The majority of the variables are \textit{BY Dra} type main-sequence stars, usually K or M, exhibiting variations due to rotation coupled with star spots and other chromospheric activity. They investigate the relations between period, color, age, and activity measures, including optical flaring, for K and M dwarfs and determined that the fraction of stars that is variable with amplitudes greater than 0.01 mag and periods between 0.1 and 100 days, increases exponentially such that $\sim$ 50 per cent  of field dwarf stars in the solar neighbourhood are variable at this level and at these timescales. 

\cite{2014arXiv1410.0014S} presented the colours and activity of ultra-cool dwarfs from the Tenth Data Release of the Sloan Digital Sky Survey (SDSS) combining the previous samples of SDSS M and L dwarfs with new data obtained from the Baryon Oscillation Sky Survey (BOSS) to produce the BOSS Ultra-cool Dwarf (BUD) sample of 11\,820 dwarfs. They obtained the fraction of active dwarfs rises through the M spectral sequence until it reaches  $\sim$90 per cent at spectral type L0, through the presence of H$\alpha$ emission. The fraction of active dwarfs then declines to 40 per cent at spectral type M5 (mass $\sim$ 0.21\,M$_{\odot}$, by the \cite{2000asqu.book..381D} classification). Considering M dwarfs only within 100\,pc from the Galactic plane, the activity fraction increases to 10 per cent at type M0 (mass $\sim$ 0.51\,M$_{\odot}$) and 50 per cent at type M4.

On the other hand, several studies of the low mass stars have been performed by the \textit{Kepler} mission - a space telescope whose principal purpose was to detect transits and to discover exoplanets, that provides an great opportunity to study the light-curves of stars with never before studied precision and coverage. In the first look at a large sample of stars with photometric data of a quality that has heretofore been only available for our Sun, \cite{2010ApJ...713L.155B} found that nearly half of their sample was more active than the active Sun.

\cite{2015arXiv151100957N} using photometry from the MEarth transit survey detected rotation periods between 0.1 and 150 days for 391 nearby mid-to-late M dwarfs in the Northern hemisphere. For fully-convective stars with detected rotation periods, they found no correlation between metallicity and rotation period or amplitude.

Estimating M dwarf rotation periods, \cite{2015ApJ...812....3W} used photometric observations from the Earth survey for transiting exoplanets. Using spectroscopic observations and photometric light curves of 238 nearby M dwarfs, they examined the relationships between magnetic activity by H$\alpha$ emission, rotation period, and stellar age. The amplitude of the photometric modulations detected were typically 0.5 to 2\% peak-to-peak. For all M spectral types, they found that the presence of magnetic activity is tied to rotation, including for late-type, fully convective M dwarfs.

\section{Variable stars}

\begin{figure}
\centering
\includegraphics[width=84mm,clip=]{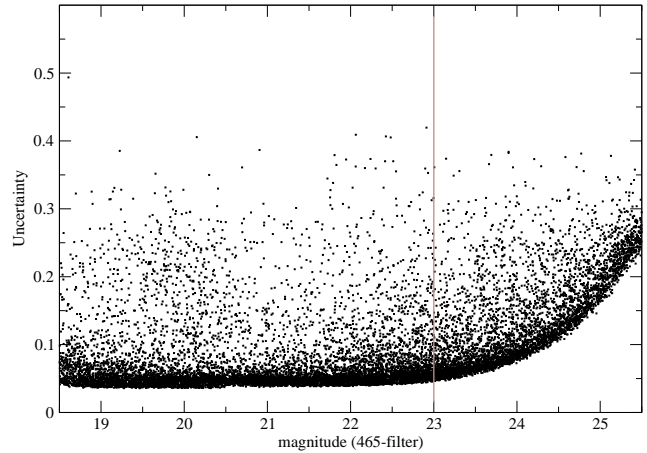}
\caption{ Uncertainty as the function of the magnitude in the 465-filter. The vertical line shows the magnitude limit used in this work. }
\label{sigma}
\end{figure}

\begin{figure}
\centering
\includegraphics[width=84mm,clip=]{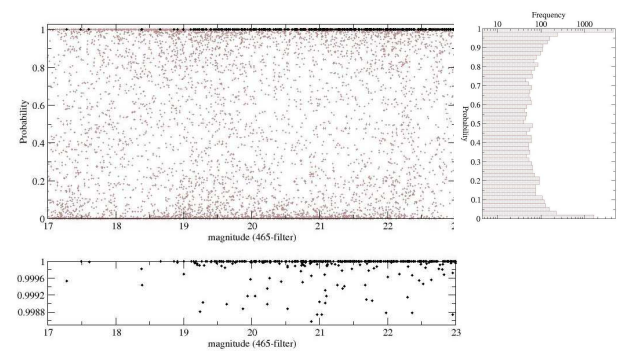} 
 \caption{  Distribution of probability of variability in the 465nm-filter of NGC\,6397 stars.
The stars selected as candidates for variables have confidence level of 99.9\%, with amplitudes variation greater than $\pm$ 0.2 magnitude. The top shows all the 9\,868 stars studied. Bottom panel shows a zoom of the region with probability above 99.9\%, containing 412 variable stars. The right panel shows frequency of the probability distribution function of the number of stars.}
\label{prob}
\end{figure}

For each light curve calculate the \textit{chi-square},
\begin{equation}
\chi^{2} = \sum_{i = 1}^N \, \left(\frac{m_i - \bar{m}}{\sigma_i} \right)^2  \quad ,
\label{chi2}
\end{equation}
where, $m_i$ and $\sigma_i$ are the $i^{th}$ magnitude measure and its uncertainty and $\bar{m}$ the average magnitude. Both stellar variability and underestimated uncertainties can lead to high values of $\chi^{2}$. On other hand, overestimated uncertainties can lead to low values for $\chi^{2}$.
The probability of a \textit{chi-square} distribution with ($N-1$) degrees of freedom have a value less than $\chi^{2}$ by chance can be calculated directly from the \textit{chi-square} cumulative distribution function.
Figure~\ref{prob} shows the calculated probabilities in function of the magnitude for the 9\,868 stars found in our NGC\,6397 images.
The Fig.~\ref{sigma} shows the uncertainty as function of the magnitude in the 465-filter, which was used in Equation~\ref{chi2}.

\begin{figure}
\centering
\includegraphics[width=84mm,clip=]{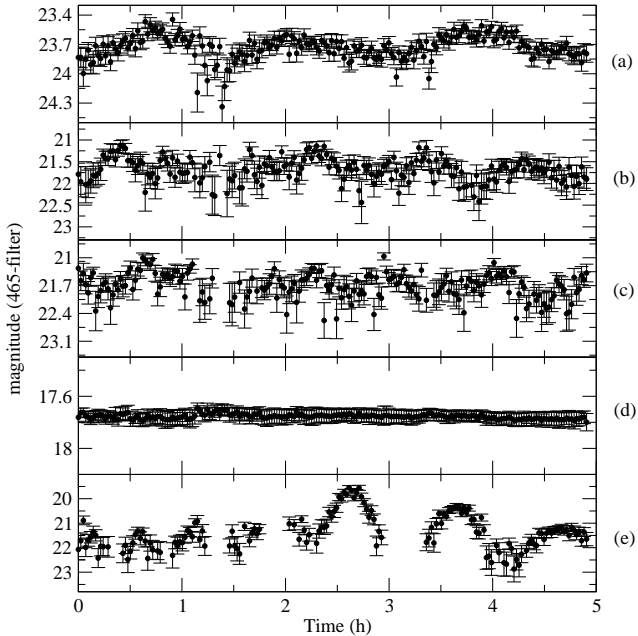}
\caption{ Panels (a), (b) and (c) are exemples of light curve selected as variable stars; (d) is a typical light curve of a normal star, while (e) is an example of light curve with contamination by diffraction rays of a neighbour star brightness, eliminated by visual inspection.  }
\label{painel}
\end{figure}

After sorting by the probability and $\chi^2$ variability method, those that showed variability in one or both the nights with amplitude variation greater than 0.2 mag were selected. 
However, it was necessary to make a visual inspection of the light curve to remove false alarms, discarding more than half of the candidates.
Examples of candidate variables are selected stars shown in Fig.~\ref{painel} (a), (b) and (c). A light curve of a star considered normal or not variable, is shown in (d) Fig.~\ref{painel}.
More than half of the stars selected as candidates for variables were false alarms due to contamination of diffraction rays of bright stars, as the example shown in (e) Fig.~\ref{painel}, or large brightness variation of the estimated background.

\begin{figure*}
\begin{minipage}{115mm}
\centering
\includegraphics[width=115mm,clip=]{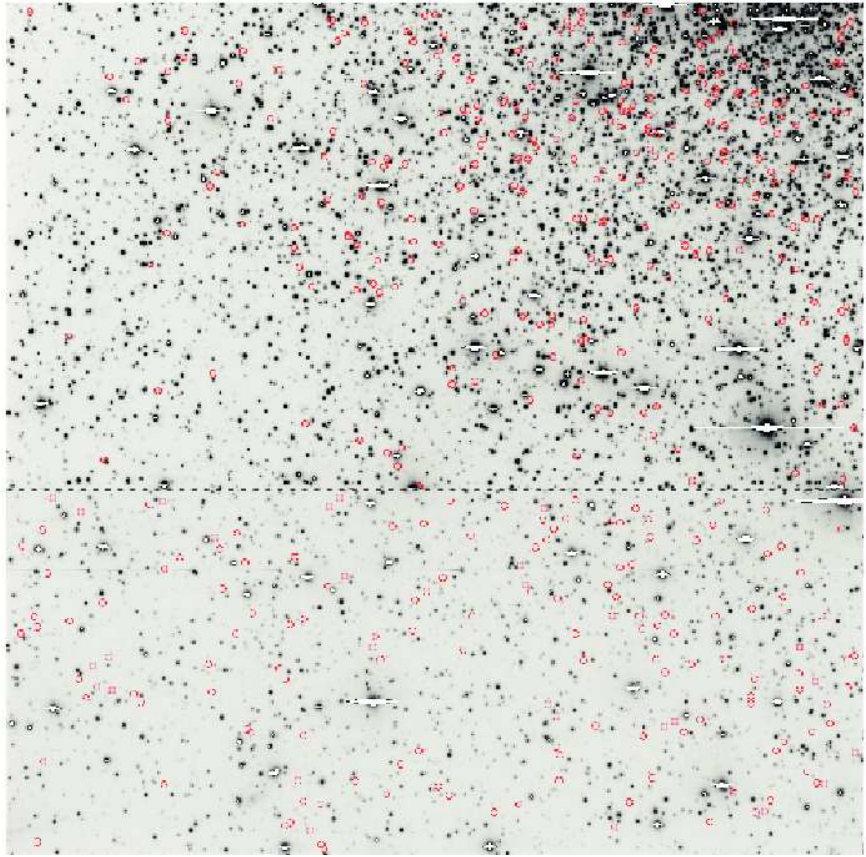}
\caption{ The circles show the positions of variables in images obtained with the ESO-VLT. Above and below CCD1, CCD2. The dashed line shows the boundary between the two CCDs. }
\label{position}
\end{minipage}
\end{figure*}

We found a total of  412  variables stars with 99.9\% confidence level, corresponding to 4.8 $\pm$ 0.2 per cent of the observed stars between between magnitude 17 and 23 in the 465nm-filter. The objects brighter than V $\simeq $ 17 were saturated, so it was not possible in this study, to search for some rapidly rotating stars, contact binaries, RR Lyrae, or SX Phe stars. Fig.~\ref{position} shows the positions of variables (circles) in images obtained with the ESO-VLT. 

\begin{figure}
\centering
\includegraphics[width=84mm,clip=]{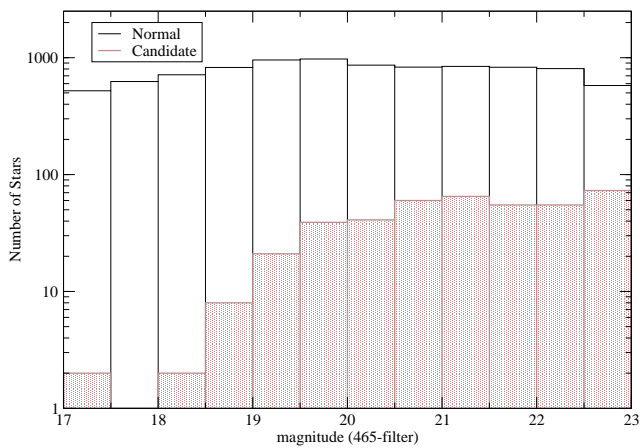}
\caption{ Magnitude distribution with the number of normal stars and magnitude with the number of variables stars candidate.}
\label{compara2}
\end{figure}

\begin{figure}
\centering
\includegraphics[width=84mm,clip=]{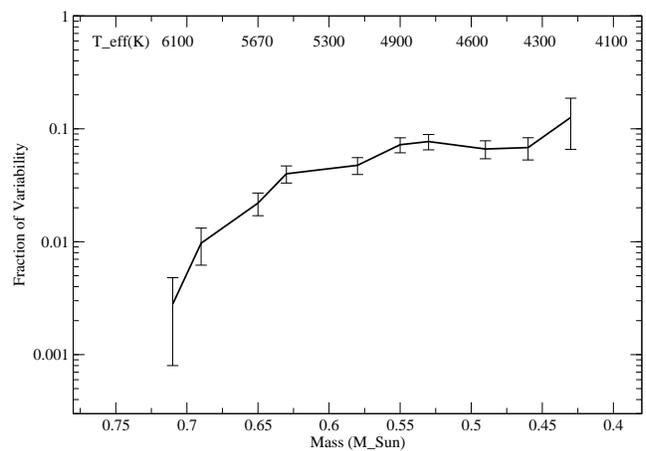}
\caption{ Fraction of candidates to variable in function of the stellar mass for NGC\,6397 with magnitude between 17 and 23 (465nm-filter).
The mass were estimated from the by isochrones calculated by A. Dotter (2008). The uncertainty bars are including the correction of uncertainty in completeness. 
}
\label{compara}
\end{figure}

Figure~\ref{compara2} shows the histogram of not variable stars and the magnitude compared with the number of variables stars candidate and Figure~\ref{compara} shows the fraction of variables as a function of the stellar mass for NGC\,6397 with magnitude between 17 and 23 (465nm-filter). 

The cold stars located in the lower main sequence commonly present high chromospheric activity and will have spots on their surfaces \citep[e.g.][]{2012IAUS..286..307V}.  
Convection is an important form of the transport of energy and momentum in stellar interiors.  For low-temperature stars with extended convective envelope, convection exceed radiation and becomes the major form of energy transportation \citep[e.g.][]{2016MNRAS.457.3163X}.
The active fraction peaks at spectral type M8 and drops sharply at late-M spectral types \citep[e.g.][] {2004AJ....128..426W}.
The coronal X-ray emission levels of L dwarfs are generally below the current sensitivity limits \citep{2006A&A...448..293S}. 
%Convection is so deep that it causes chromospherical variability. 
As the convection layer is deepening with the decreasing effective temperature  the activity fraction of variability increase when the mass of stars decrease is expected.

\section{Discussion and conclusion}

We have conducted a photometric survey of the globular cluster NGC\,6397 in a search for variable stars.
We analysed 8\,391 light curves of NGC\,6397 stars which lie in the lower part of the main-sequence in the colour-magnitude diagram, between magnitude 17 and 23 (465nm-filter) where are stars of spectral types K and M. We identified 412 stars, reaching $\sim$ 4.8 $\pm$ 0.2 per cent of variability with timescales between 0.004 and 2\,d with 99.9\% confidence level of variability to be real.

The relationship between the fraction of candidates to variables (with amplitudes variation greater than $\pm$ 0.2 magnitude) in function of the stellar mass of the NGC\,6397 stars is shown in Fig.~\ref{compara}. From the curves it is observed that the fraction of the variation increases as the mass decreases. \cite{2011AJ....141..166H} also found the increased variability with decreased mass of the studied field stars. However, the stars that we observe are of low metallicity.

\section*{Acknowledgments}

We thank the referee for important comments and suggestions.
We thanks to ESO-VLT data program ID, support team. 
We acknowledge financial support from the Brazilian Institution CNPq and CAPES.

\section{Appendix}

The stars identified (ID) is compared by the CCD number (1 or 2) and the ID provides by IRAF, the coordinates (R.A. and Dec. for J2000), reduced Chi-squere [$ \chi^2$ red = $ \chi^2$/(N-1)], the average magnitude in the B$\_HIGH$, V$\_HIGH$ and FILT\_465\_250 filters of the identified variable stars in NGC\,6397 are listed in Table~\ref{Mvariables}. Some stars we did not get the magnitudes of some filters.

\begin{table*}
\begin{minipage}{180mm}
\caption[]{Variable stars in NGC\,6397 with 99.9\% confidence level. } 
    \label{Mvariables}
    \vspace{1em}
    \centering
\begin{tabular}{ c c c c c c c  c c c} 
\hline\hline
Star &  R.A. (J2000)    &  Dec. (J2000) &  $\chi^2_{\rm{red}}$  & $\langle$FILT\_465\_250$\rangle$ &  $\langle$B$\_HIGH \rangle$ & $\sigma_B$ &  $\langle$V$\_HIGH \rangle$ &  $\sigma_V$    \\
 ID   & \textit{h m s} &    ${}^{\circ} $ '  "  	  &              &         &		  &       &		         &           \\
\hline  
1/	103	&	17:40:47.42  & -53:45:11.61	&	1.59		&	22.81	&	23.56	&	0.05	&	22.05	&	0.03	\\
1/	114	&	17:41:02.70  & -53:45:12.39	&	2.59		&	21.79	&	16.47	&	0.18	&	16.77	&	0.19	\\
1/	122	&	17:40:41.69  & -53:45:07.80	&	2.57		&	21.37	&	21.37	&	0.07	&	20.43	&	0.08	\\
1/	224	&	17:40:41.69  & -53:45:07.89	&	1.79		&	21.42	&	21.81	&	0.04	&	20.64	&	0.04	\\
1/	322	&	17:41:03.84  & -53:45:05.49	&	1.94		&	22.48	&	22.61	&	0.08	&	21.31	&	0.06	\\
1/	515	&	17:40:53.86  & -53:44:59.11	&	1.55		&	19.53	&	19.81	&	0.01	&	19.01	&	0.01	\\
1/	571	&	17:41:04.40  & -53:44:58.59 &	2.49		&	22.32	&	22.73	&	0.08	&	21.20	&	0.04	\\
1/	651	&	17:41:03.41  & -53:44:56.71	&	1.70		&	21.70	&	21.81	&	0.07	&	20.52	&	0.10	\\
1/	654	&	17:40:48.13  & -53:44:56.57	&	2.32		&	22.71	&	23.15	&	0.05	&	21.83	&	0.03	\\
1/	814	&	17:41:05.67  & -53:44:51.69	&	1.63		&	22.23	&	21.57	&	0.05	&	20.69	&	0.04	\\
1/	875	&	17:40:49.61  & -53:44:49.05	&	1.40		&	23.00	&	23.63	&	0.04	&	22.07	&	0.02	\\
1/	1021	&	17:41:08.85  & -53:44:43.53	&	1.54		&	23.00	&	23.71	&	0.06	&	22.11	&	0.03	\\
1/	1204	&	17:40:52.02  & -53:44:37.78	&	2.01 	&	22.62	&	23.21	&	0.06	&	21.99	&	0.05	\\
1/	1214	&	17:40:50.11  & -53:44:37.77	&	1.92		&	22.68	&	23.86	&	0.08	&	22.16	&	0.04	\\
1/	1266	&	17:40:52.65  & -53:44:36.53	&	2.07		&	22.96	&	23.10	&	0.04	&	21.81	&	0.03	\\
1/	1305	&	17:40:49.47  & -53:44:34.63	&	1.96		&	22.71	&	23.30	&	0.07	&	21.67	&	0.04	\\
1/	1347	&	17:40:39.08  & -53:45:10.90	&	1.97		&	21.74	&	21.77	&	0.06	&	20.48	&	0.06	\\
1/	1357	&	17:40:40.17  & -53:44:34.02	&	1.40		&	19.16	&	19.51	&	0.01	&	18.74	&	0.02	\\
1/	1416	&	17:40:56.29  & -53:44:31.87	&	2.29		&	22.19	&	22.29	&	0.08	&	20.90	&	0.06	\\
1/	1576	&	17:41:17.07  & -53:44:25.80	&	1.83		&	22.99	&	24.16	&	0.07	&	22.58	&	0.04	\\
1/	1672	&	17:41:00.87  & -53:44:24.35	&	1.35		&	22.36	&	22.90	&	0.04	&	21.37	&	0.03	\\
1/	1715	&	17:40:59.45  & -53:44:22.85	&	1.44		&	21.62	&	21.96	&	0.02	&	20.82	&	0.03	\\
1/	1740	&	17:40:53.66  & -53:44:22.08	&	1.89		&	22.12	&	22.67	&	0.05	&	21.26	&	0.03	\\
1/	1839	&	17:41:14.10  & -53:44:20.55	&	2.37		&	22.66	&	24.69	&	0.14	&	22.98	&	0.08	\\
1/	1905	&	17:40:40.51  & -53:44:15.96	&	1.53		&	22.06	&	23.13	&	0.07	&	21.81	&	0.07	\\
1/	1941	&	17:41:00.70  & -53:44:16.08	&	2.23		&	22.45	&	23.27	&	0.09	&	21.88	&	0.06	\\
1/	2111	&	17:40:58.49  & -53:44:10.81	&	2.84		&	22.96	&	20.73	&	0.10	&	19.83	&	0.24	\\
1/	2136	&	17:40:55.18  & -53:44:10.80	&	1.76		&	23.00	&	23.57	&	0.05	&	22.33	&	0.05	\\
1/	2209	&	17:40:51.36  & -53:44:08.32	&	1.85		&	22.74	&	23.09	&	0.05	&	21.72	&	0.02	\\
1/	2385	&	17:40:53.61  & -53:44:02.89	&	1.86		&	20.36	&	21.18	&	0.08	&	20.08	&	0.12	\\
1/	2387	&	17:40:38.89  & -53:44:01.56	&	1.30		&	20.96	&	21.36	&	0.02	&	20.27	&	0.03	\\
1/	2436	&	17:40:53.41  & -53:44:01.38	&	2.19		&	22.94	&	21.97	&	0.07	&	20.48	&	0.06	\\
1/	2441	&	17:40:38.89  & -53:44:02.77	&	1.71		&	21.11	&	21.49	&	0.02	&	20.36	&	0.03	\\
1/	2687	&	17:40:55.22  & -53:43:54.73	&	1.75		&	22.69	&	22.84	&	0.05	&	21.33	&	0.03	\\
1/	2694	&	17:40:55.93  & -53:43:54.43	&	1.89		&	22.57	&	23.29	&	0.06	&	21.65	&	0.02	\\
1/	2721	&	17:40:41.21  & -53:43:53.72	&	1.49		&	22.71	&	23.39	&	0.07	&	21.94	&	0.03	\\
1/	2752	&	17:40:42.40  & -53:43:52.83	&	1.55		&	22.04	&	22.41	&	0.04	&	21.11	&	0.03	\\
1/	2782	&	17:40:53.58  & -53:43:52.30	&	1.32		&	21.10	&	21.91	&	0.06	&	20.60	&	0.04	\\
1/	2804	&	17:40:46.53  & -53:43:51.96	&	1.38		&	21.84	&	22.62	&	0.04	&	21.10	&	0.03	\\
1/	2836	&	17:40:56.85  & -53:43:50.50	&	2.17		&	22.35	&	22.59	&	0.04	&	21.14	&	0.03	\\
1/	2846	&	17:40:55.18  & -53:43:50.80	&	2.63		&	22.45	&	23.06	&	0.12	&	21.68	&	0.10	\\
1/	2894	&	17:40:58.96  & -53:43:49.61	&	1.56		&	22.89	&	23.36	&	0.06	&	22.05	&	0.06	\\
1/	2923	&	17:40:39.17  & -53:43:48.23	&	1.79		&	22.71	&	23.51	&	0.03	&	22.02	&	0.02	\\
1/	3002	&	17:40:50.55  & -53:43:46.45	&	2.03		&	20.40	&	20.76	&	0.01	&	19.78	&	0.01	\\
1/	3019	&	17:40:38.89  & -53:43:45.72	&	2.06		&	21.45	&	21.63	&	0.03	&	20.51	&	0.03	\\
1/	3130	&	17:40:40.66  & -53:43:43.23	&	2.07		&	22.29	&	22.79	&	0.12	&	21.42	&	0.02	\\
1/	3157	&	17:41:00.16  & -53:43:42.72	&	2.75		&	22.28	&	23.25	&	0.09	&	21.72	&	0.05	\\
1/	3312	&	17:41:04.75  & -53:43:38.32	&	1.83		&	21.85	&	22.07	&	0.03	&	20.80	&	0.02	\\
1/	3346	&	17:41:06.10  & -53:43:37.69	&	1.99		&	22.90	&	23.40	&	0.07	&	21.92	&	0.03	\\
1/	3402	&	17:41:08.57  & -53:43:35.81	&	1.35		&	22.36	&	22.80	&	0.03	&	21.68	&	0.02	\\
1/	3404	&	17:41:03.98  & -53:43:35.82	&	1.36		&	22.75	&	23.55	&	0.06	&	21.98	&	0.02	\\
1/	3424	&	17:40:40.87  & -53:43:35.70	&	1.34		&	22.46	&	22.96	&	0.05	&	21.53	&	0.03	\\
1/	3586	&	17:40:48.78  & -53:43:31.39	&	2.42		&	19.49	&	20.31	&	0.04	&	19.29	&	0.04	\\
1/	3756	&	17:40:41.65  & -53:43:26.93	&	1.59		&	22.22	&	22.50	&	0.03	&	21.14	&	0.02	\\
1/	3791	&	17:41:09.27  & -53:43:25.77	&	1.50		&	22.94	&	23.63	&	0.05	&	22.07	&	0.02	\\
1/	3797	&	17:40:49.70  & -53:43:25.75	&	1.30		&	20.88	&	21.13	&	0.01	&	20.06	&	0.02	\\
1/	3841	&	17:40:52.39  & -53:43:24.51	&	1.66		&	21.16	&	21.42	&	0.02	&	20.29	&	0.03	\\
1/	3847	&	17:41:17.47  & -53:43:23.84	&	2.15		&	19.99	&	20.34	&	0.02	&	19.56	&	0.03	\\
1/	3899	&	17:41:03.06  & -53:43:23.28	&	2.54		&	21.50s	&	21.47	&	0.03	&	20.20	&	0.04	\\
1/	3908	&	17:41:04.89  & -53:43:22.65	&	1.30		&	22.96	&	23.66	&	0.08	&	22.09	&	0.02	\\
1/	3950	&	17:40:53.09  & -53:43:22.00	&	1.52		&	22.05	&	22.37	&	0.03	&	20.96	&	0.02	\\
\hline
\end{tabular} 
\end{minipage}
\end{table*}

\begin{table*}
\begin{minipage}{180mm}
\caption[]{Continuation. } 
    \label{}
    \vspace{1em}
    \centering
\begin{tabular}{ c c c c c c c  c c c} 
\hline\hline
Star &  R.A. (J2000)    &  Dec. (J2000) &  $\chi^2_{\rm{red}}$  & $\langle$FILT\_465\_250$\rangle$ &  $\langle$B$\_HIGH \rangle$ & $\sigma_B$ &  $\langle$V$\_HIGH \rangle$ &  $\sigma_V$    \\
 ID   & \textit{h m s} &    ${}^{\circ} $ '  "  	  &              &         &		  &       &		         &           \\
\hline 
1/	3993	&	17:40:57.40  & -53:43:20.76	&	2.46		&	22.84	&	23.22	&	0.06	&	21.68	&	0.03	\\
1/	3994	&	17:40:52.18  & -53:43:21.37	&	1.91		&	22.24	&	22.59	&	0.11	&	21.27	&	0.10	\\
1/	4001	&	17:40:42.43  & -53:43:19.42	&	1.73		&	19.45	&	20.50	&	0.04	&	19.72	&	0.13	\\
1/	4041	&	17:40:47.23  & -53:43:19.46	&	2.51		&	21.36	&	22.16	&	0.05	&	20.95	&	0.07	\\
1/	4093	&	17:40:57.26  & -53:43:18.26	&	1.39		&	19.20	&	19.57	&	0.01	&	18.80	&	0.01	\\
1/	4111	&	17:40:46.45  & -53:43:18.20	&	1.95		&	20.65	&	21.29	&	0.05	&	20.33	&	0.07	\\
1/	4119	&	17:40:44.76  & -53:43:18.18	&	2.73		&	22.98	&	22.08	&	0.07	&	20.22	&	0.12	\\
1/	4152	&	17:41:09.56  & -53:43:16.99	&	2.63		&	22.50	&	22.50	&	0.03	&	21.07	&	0.03	\\
1/	4187	&	17:40:47.87  & -53:43:15.70	&	1.83		&	22.98	&	23.68	&	0.06	&	22.08	&	0.03	\\
1/	4205	&	17:41:06.02  & -53:43:15.12	&	2.29		&	20.63	&	20.44	&	0.17	&	19.52	&	0.03	\\
1/	4215	&	17:41:01.43  & -53:43:14.50	&	1.76		&	21.75	&	22.72	&	0.06	&	21.45	&	0.06	\\
1/	4374	&	17:41:06.59  & -53:43:09.48	&	1.61		&	22.98	&	23.46	&	0.11	&	21.99	&	0.08	\\
1/	4414	&	17:41:16.62  & -53:43:08.80	&	2.77		&	22.68	&	22.54	&	0.08	&	21.31	&	0.07	\\
1/	4508	&	17:41:09.63  & -53:43:06.96	&	1.42		&	21.58	&	21.97	&	0.03	&	20.72	&	0.03	\\
1/	4523	&	17:40:50.06  & -53:43:06.94	&	2.01		&	20.72	&	20.77	&	0.02	&	19.59	&	0.02	\\
1/	4543	&	17:40:41.79  & -53:43:06.24	&	1.51		&	22.87	&	23.56	&	0.05	&	22.00	&	0.03	\\
1/	4561	&	17:41:12.45  & -53:43:05.06	&	2.04		&	22.63	&	22.86	&	0.04	&	21.51	&	0.03	\\
1/	4633	&	17:40:52.39  & -53:43:03.81	&	1.86		&	20.55	&	20.83	&	0.03	&	19.86	&	0.03	\\
1/	4681	&	17:41:03.06  & -53:43:02.58	&	1.35		&	22.96	&	23.61	&	0.03	&	22.03	&	0.02	\\
1/	4691	&	17:40:41.16  & -53:43:02.47	&	1.69		&	21.80	&	22.17	&	0.05	&	20.97	&	0.05	\\
1/	4692	&	17:40:40.45  & -53:43:01.84	&	1.55		&	21.71	&	22.41	&	0.04	&	21.09	&	0.02	\\
1/	4702	&	17:40:53.80  & -53:43:01.94	&	2.47		&	22.90	&	23.88	&	0.09	&	22.24	&	0.04	\\
1/	4703	&	17:40:53.45  & -53:43:01.94	&	1.35		&	22.83	&	22.80	&	0.05	&	21.29	&	0.04	\\
1/	4767	&	17:40:42.99  & -53:42:59.98	&	2.40		&	22.53	&	23.14	&	0.05	&	21.92	&	0.04	\\
1/	4830	&	17:40:59.52  & -53:42:58.82	&	1.75		&	20.26	&	20.60	&	0.06	&	19.58	&	0.06	\\
1/	4952	&	17:40:48.29  & -53:42:56.27	&	1.86		&	22.10	&	22.47	&	0.04	&	21.36	&	0.03	\\
1/	4966	&	17:40:45.89  & -53:42:55.62	&	2.45		&	22.66	&	23.36	&	0.14	&	22.00	&	0.09	\\
1/	4972	&	17:40:45.18  & -53:42:54.99	&	1.86		&	21.80	&	21.37	&	0.07	&	20.24	&	0.10	\\
1/	5020	&	17:41:04.68  & -53:42:54.43	&	2.23		&	21.91	&	21.76	&	0.07	&	20.65	&	0.04	\\
1/	5094	&	17:41:07.65  & -53:42:51.91	&	1.37		&	21.80	&	22.15	&	0.06	&	20.92	&	0.05	\\
1/	5127	&	17:40:43.84  & -53:42:51.84	&	2.17		&	22.64	&	22.79	&	0.08	&	21.23	&	0.07	\\
1/	5155	&	17:41:05.74  & -53:42:51.29	&	1.41		&	19.76	&	19.88	&	0.01	&	19.04	&	0.01	\\
1/	5247	&	17:40:56.84  & -53:42:50.66	&	1.37		&	22.30	&	22.88	&	0.05	&	21.47	&	0.04	\\
1/	5250	&	17:40:38.33  & -53:42:49.27	&	1.49		&	22.69	&	23.64	&	0.06	&	22.04	&	0.03	\\
1/	5361	&	17:41:00.30  & -53:42:46.28	&	1.33		&	22.52	&	22.96	&	0.04	&	21.59	&	0.02	\\
1/	5367	&	17:41:14.22  & -53:42:46.24	&	1.58		&	22.81	&	22.94	&	0.06	&	21.65	&	0.03	\\
1/	5383	&	17:40:57.41  & -53:42:45.65	&	2.28		&	22.99	&	22.98	&	0.05	&	21.39	&	0.04	\\
1/	5452	&	17:40:42.85  & -53:42:44.30	&	3.10		&	19.47	&	20.08	&	0.07	&	19.14	&	0.07	\\
1/	5492	&	17:40:44.69  & -53:42:43.07	&	1.82		&	22.39	&	22.26	&	0.05	&	21.26	&	0.08	\\
1/	5508	&	17:40:44.62  & -53:42:43.07	&	1.56		&	21.27	&	21.45	&	0.04	&	20.33	&	0.03	\\
1/	5530	&	17:40:38.62  & -53:42:43.00	&	1.39		&	20.86	&	21.24	&	0.02	&	20.15	&	0.01	\\
1/	5537	&	17:40:52.81  & -53:42:43.12	&	1.90		&	22.30	&	22.72	&	0.05	&	21.42	&	0.04	\\
1/	5573	&	17:40:51.33  & -53:42:41.23	&	1.65		&	19.41	&	19.84	&	0.02	&	19.02	&	0.02	\\
1/	5622	&	17:40:39.89  & -53:42:40.51	&	1.39		&	19.20	&	19.34	&	0.01	&	18.62	&	0.01	\\
1/	5663	&	17:41:13.79  & -53:42:39.34	&	2.12		&	22.46	&	22.76	&	0.03	&	21.32	&	0.02	\\
1/	5669	&	17:40:38.90  & -53:42:38.62	&	2.22		&	21.06	&	22.00	&	0.06	&	20.76	&	0.07	\\
1/	5691	&	17:40:43.07  & -53:42:38.66	&	1.31		&	21.77	&	22.50	&	0.07	&	21.30	&	0.05	\\
1/	5705	&	17:40:52.39  & -53:42:39.36	&	1.54		&	20.38	&	20.88	&	0.03	&	19.84	&	0.02	\\
1/	5757	&	17:41:00.16  & -53:42:37.50	&	2.28		&	19.45	&	19.94	&	0.03	&	19.17	&	0.11	\\
1/	5784	&	17:40:41.94  & -53:42:36.77	&	2.43		&	22.47	&	22.55	&	0.05	&	21.01	&	0.03	\\
1/	5793	&	17:41:04.61  & -53:42:36.24	&	2.12		&	21.27	&	22.25	&	0.04	&	21.12	&	0.04	\\
1/	5833	&	17:41:03.34  & -53:42:35.62	&	1.74		&	22.68	&	23.43	&	0.07	&	21.85	&	0.02	\\
1/	5834	&	17:40:59.45  & -53:42:35.62	&	1.67		&	22.09	&	22.68	&	0.04	&	21.23	&	0.02	\\
1/	5876	&	17:41:05.18  & -53:42:34.36	&	1.41		&	22.05	&	22.42	&	0.05	&	21.08	&	0.04	\\
1/	5893	&	17:40:45.54  & -53:42:34.30	&	1.33		&	21.67	&	22.17	&	0.02	&	21.04	&	0.02	\\
1/	5921	&	17:40:46.60  & -53:42:33.68	&	2.06		&	22.19	&	22.46	&	0.03	&	21.10	&	0.03	\\
1/	5947	&	17:40:56.56  & -53:42:33.10	&	1.55		&	20.87	&	21.45	&	0.05	&	20.32	&	0.03	\\
1/	5970	&	17:41:04.40  & -53:42:33.11	&	1.44		&	20.06	&	20.92	&	0.03	&	19.92	&	0.02	\\
1/	5971	&	17:40:57.12  & -53:42:32.48	&	1.68		&	20.40	&	20.59	&	0.02	&	19.46	&	0.02	\\
1/	5972	&	17:40:54.02  & -53:42:32.47	&	1.50		&	22.92	&	23.65	&	0.04	&	22.10	&	0.02	\\
1/	5986	&	17:40:56.14  & -53:42:26.83	&	1.58		&	22.74	&	23.22	&	0.07	&	21.87	&	0.04	\\
\hline
\end{tabular} 
\end{minipage}
\end{table*}

\begin{table*}
\begin{minipage}{180mm}
\caption[]{Continuation. } 
    \label{}
    \vspace{1em}
    \centering
\begin{tabular}{ c c c c c c c  c c c} 
\hline\hline
Star &  R.A. (J2000)    &  Dec. (J2000) &  $\chi^2_{\rm{red}}$  & $\langle$FILT\_465\_250$\rangle$ &  $\langle$B$\_HIGH \rangle$ & $\sigma_B$ &  $\langle$V$\_HIGH \rangle$ &  $\sigma_V$    \\
 ID   & \textit{h m s} &    ${}^{\circ} $ '  "  	  &              &         &		  &       &		         &           \\
\hline
1/	5987	&	17:40:43.63  & -53:42:32.40	&	1.42		&	22.61	&	22.93	&	0.08	&	21.76	&	0.07	\\
1/	6005	&	17:41:07.86  & -53:42:31.22	&	2.14		&	22.87	&	23.75	&	0.07	&	22.27	&	0.03	\\
1/	6033	&	17:40:49.50  & -53:42:31.19	&	1.49		&	22.13	&	22.70	&	0.05	&	21.45	&	0.03	\\
1/	6077	&	17:40:45.68  & -53:42:29.28	&	1.32		&	19.94	&	19.90	&	0.02	&	19.08	&	0.04	\\
1/	6086	&	17:40:50.56  & -53:42:29.32	&	1.58		&	21.73	&	21.94	&	0.03	&	20.85	&	0.02	\\
1/	6111	&	17:40:45.75  & -53:42:29.91	&	1.44		&	19.40	&	19.69	&	0.02	&	18.89	&	0.04	\\
1/	6177	&	17:40:49.78  & -53:42:28.06 &	1.63		&	22.33	&	23.57	&	0.20	&	21.85	&	0.05	\\
1/	6214	&	17:40:56.06  & -53:42:26.83	&	2.65		&	22.08	&	22.42	&	0.04	&	21.10	&	0.02	\\
1/	6243	&	17:40:59.17  & -53:42:26.21	&	1.40		&	22.93	&	23.70	&	0.05	&	22.11	&	0.02	\\
1/	6341	&	17:40:56.28  & -53:42:23.70	&	1.55		&	21.80	&	22.72	&	0.10	&	21.34	&	0.06	\\
1/	6363	&	17:40:49.28  & -53:42:23.66	&	2.19		&	22.85	&	18.86	&	0.29	&	18.32	&	0.15	\\
1/	6389	&	17:40:58.96  & -53:42:23.08	&	2.05 	&	22.40	&	22.62	&	0.25	&	21.46	&	0.02	\\
1/	6399	&	17:40:55.01  & -53:42:22.44	&	1.34		&	22.00	&	21.99	&	0.02	&	21.02	&	0.03	\\
1/	6467	&	17:40:57.48  & -53:42:21.19	&	1.57		&	22.05	&	21.34	&	0.09	&	20.27	&	0.10	\\
1/	6523	&	17:40:42.79  & -53:42:20.48	&	1.54		&	22.69	&	23.55	&	0.06	&	22.02	&	0.06	\\
1/	6529	&	17:41:01.15  & -53:42:19.94	&	1.35		&	22.17	&	22.76	&	0.03	&	21.37	&	0.02	\\
1/	6550	&	17:41:07.36  & -53:42:18.68 &	1.57		&	22.52	&	23.23	&	0.05	&	21.92	&	0.03	\\
1/	6658	&	17:40:49.78  & -53:42:16.77	&	2.31		&	21.59	&	21.54	&	0.08	&	20.36	&	0.06	\\
1/	6679	&	17:40:45.68  & -53:42:17.37	&	1.95		&	20.21	&	20.75	&	0.03	&	19.71	&	0.02	\\
1/	6688	&	17:40:48.51  & -53:42:16.13	&	2.13		&	22.31	&	22.50	&	0.08	&	21.41	&	0.11	\\
1/	6691	&	17:40:41.66  & -53:42:16.07	&	1.88		&	22.84	&	23.61	&	0.16	&	22.24	&	0.09	\\
1/	6764	&	17:40:46.39  & -53:42:14.24	&	1.70		&	19.90	&	20.00	&	0.01	&	19.15	&	0.01	\\
1/	6775	&	17:40:51.97  & -53:42:14.90	&	1.67		&	20.91	&	21.02	&	0.02	&	19.91	&	0.03	\\
1/	6776	&	17:40:51.69  & -53:42:14.27	&	2.03		&	22.88	&	23.53	&	0.06	&	22.04	&	0.06	\\
1/	6778	&	17:40:43.42  & -53:42:14.21	&	2.58		&	22.72	&	22.97	&	0.07	&	21.51	&	0.04	\\
1/	6791	&	17:41:16.55  & -53:42:13.61	&	2.86		&	22.68	&	22.97	&	0.05	&	21.47	&	0.02	\\
1/	6812	&	17:40:59.31  & -53:42:13.67	&	1.43		&	20.60	&	21.06	&	0.02	&	20.03	&	0.03	\\
1/	6848	&	17:41:10.82  & -53:42:12.40	&	2.02		&	22.90	&	23.29	&	0.10	&	21.71	&	0.07	\\
1/	6866	&	17:40:56.77  & -53:42:13.04	&	2.45		&	20.91	&	21.60	&	0.04	&	20.47	&	0.03	\\
1/	6891	&	17:40:42.44  & -53:42:11.69	&	1.39 	&	21.12	&	21.52	&	0.03	&	20.53	&	0.03	\\
1/	6913	&	17:40:49.85  & -53:42:11.13	&	1.47		&	21.38	&	21.69	&	0.03	&	20.55	&	0.03	\\
1/	6931	&	17:40:40.74  & -53:42:11.05	&	2.31		&	21.90	&	22.54	&	0.07	&	21.01	&	0.08	\\
1/	6946	&	17:40:38.90  & -53:42:10.40	&	1.57		&	21.71	&	22.78	&	0.12	&	21.27	&	0.04	\\
1/	6987	&	17:40:51.33  & -53:42:09.25	&	1.32		&	22.29	&	22.62	&	0.04	&	21.18	&	0.04	\\
1/	7001	&	17:40:52.46  & -53:42:09.26	&	1.74		&	22.65	&	22.82	&	0.04	&	21.35	&	0.04	\\
1/	7018	&	17:40:39.54  & -53:42:09.15	&	1.39		&	20.96	&	21.79	&	0.04	&	20.60	&	0.03	\\
1/	7028	&	17:40:59.81  & -53:42:08.65	&	2.14		&	22.15	&	22.45	&	0.03	&	21.14	&	0.04	\\
1/	7052	&	17:41:00.09  & -53:42:08.03	&	1.36		&	21.72	&	21.80	&	0.08	&	20.51	&	0.08	\\
1/	7111	&	17:40:59.38  & -53:42:06.77	&	1.39		&	22.49	&	22.76	&	0.05	&	21.47	&	0.04	\\
1/	7112	&	17:40:57.20  & -53:42:06.77	&	2.17		&	22.29	&	22.25	&	0.08	&	21.04	&	0.08	\\
1/	7140	&	17:40:50.77  & -53:42:06.74	&	2.55		&	20.63	&	21.04	&	0.03	&	20.00	&	0.01	\\
1/	7151	&	17:40:43.92  & -53:42:06.06	&	2.05		&	21.06	&	21.23	&	0.01	&	20.13	&	0.01	\\
1/	7164	&	17:40:42.79  & -53:42:06.05	&	1.56		&	21.37	&	21.65	&	0.03	&	20.42	&	0.02	\\
1/	7175	&	17:40:46.56  & -53:42:05.31	&	1.76		&	21.02	&	21.60	&	0.02	&	20.40	&	0.02	\\
1/	7218	&	17:40:53.88  & -53:42:04.45	&	1.45		&	21.15	&	21.41	&	0.07	&	20.15	&	0.05	\\
1/	7224	&	17:40:42.37  & -53:42:04.36	&	1.36		&	20.72	&	20.83	&	0.02	&	19.67	&	0.02	\\
1/	7229	&	17:40:57.94  & -53:42:03.98	&	2.53		&	22.82	&	22.82	&	0.09	&	21.23	&	0.05	\\
1/	7238	&	17:41:18.80  & -53:42:03.56	&	1.73		&	21.89	&	22.47	&	0.05	&	21.07	&	0.04	\\
1/	7243	&	17:40:51.41  & -53:42:03.95	&	2.12		&	22.71	&	22.55	&	0.08	&	21.03	&	0.05	\\
1/	7281	&	17:40:49.73  & -53:42:03.21	&	1.90		&	21.27	&	21.35	&	0.08	&	19.99	&	0.08	\\
1/	7378	&	17:40:43.97  & -53:42:01.05	&	2.43		&	22.38	&	23.10	&	0.12	&	21.75	&	0.14	\\
1/	7392	&	17:40:46.12  & -53:42:00.26	&	2.26		&	21.88	&	21.95	&	0.02	&	20.61	&	0.02	\\
1/	7407	&	17:40:44.89  & -53:42:00.15	&	2.01		&	19.86	&	20.19	&	0.04	&	19.42	&	0.08	\\
1/	7466	&	17:40:44.04  & -53:41:58.94	&	2.55		&	22.48	&	22.73	&	0.06	&	21.65	&	0.07	\\
1/	7486	&	17:40:47.79  & -53:41:58.97	&	2.18		&	21.27	&	21.55	&	0.02	&	20.25	&	0.04	\\
1/	7501	&	17:40:46.08  & -53:42:00.47	&	2.74 	&	20.96	&	22.11	&	0.06	&	20.63	&	0.06	\\
1/	7536	&	17:40:56.27  & -53:41:57.20	&	1.61		&	20.95	&	21.21	&	0.01	&	20.04	&	0.02	\\
1/	7548	&	17:40:41.11  & -53:41:57.08	&	1.63		&	20.74	&	20.74	&	0.02	&	19.87	&	0.02	\\
1/	7587	&	17:40:46.82  & -53:41:55.56	&	1.63		&	22.65	&	22.70	&	0.08	&	21.10	&	0.06	\\
1/	7590	&	17:41:12.65  & -53:41:55.59	&	1.34		&	22.64	&	22.43	&	0.05	&	21.06	&	0.06	\\
1/	7593	&	17:41:05.65  & -53:41:56.14	&	2.18		&	21.41	&	21.61	&	0.04	&	20.44	&	0.04	\\
\hline
\end{tabular} 
\end{minipage}
\end{table*}

\begin{table*}
\begin{minipage}{180mm}
\caption[]{Continuation. } 
    \label{}
    \vspace{1em}
    \centering
\begin{tabular}{ c c c c c c c  c c c} 
\hline\hline
Star &  R.A. (J2000)    &  Dec. (J2000) &  $\chi^2_{\rm{red}}$  & $\langle$FILT\_465\_250$\rangle$ &  $\langle$B$\_HIGH \rangle$ & $\sigma_B$ &  $\langle$V$\_HIGH \rangle$ &  $\sigma_V$    \\
 ID   & \textit{h m s} &    ${}^{\circ} $ '  "  	  &              &         &		  &       &		         &           \\
\hline  
1/	7630	&	17:40:51.24  & -53:41:55.59	&	2.17		&	21.75	&	21.66	&	0.04	&	20.46	&	0.05	\\
1/	7644	&	17:40:51.00  & -53:41:53.50	&	1.51		&	21.28	&	22.03	&	0.06	&	20.81	&	0.05	\\
1/	7732	&	17:40:59.67  & -53:41:52.48	&	1.39		&	22.57	&	23.27	&	0.06	&	21.74	&	0.03	\\
1/	7743	&	17:41:19.76  & -53:41:51.62	&	1.37		&	22.56	&	23.17	&	0.05	&	21.74	&	0.04	\\
1/	7751	&	17:40:55.08  & -53:41:51.85	&	2.05		&	19.89	&	20.68	&	0.03	&	19.73	&	0.03	\\
1/	7782	&	17:40:46.05  & -53:41:51.79	&	2.07		&	22.02	&	22.19	&	0.07	&	20.85	&	0.04	\\
1/	7811	&	17:41:01.44  & -53:41:51.23	&	1.51		&	19.40	&	19.50	&	0.01	&	18.70	&	0.01	\\
1/	7813	&	17:40:45.83  & -53:41:50.54	&	1.67		&	21.58	&	22.00	&	0.03	&	20.74	&	0.02	\\
1/	7854	&	17:41:16.55  & -53:41:48.67	&	2.22	 	&	23.00	&	23.35	&	0.09	&	21.75	&	0.06	\\
1/	7881	&	17:41:10.27  & -53:41:49.33	&	1.57	  	&	22.97	&	23.48	&	0.03	&	22.00	&	0.01	\\
1/	7892	&	17:40:46.96  & -53:41:48.04	&	1.94		&	20.64	&	20.67	&	0.01	&	19.63	&	0.01	\\
1/	7967	&	17:40:44.63  & -53:41:46.14	&	1.39		&	22.81	&	22.50	&	0.12	&	21.78	&	0.06	\\
1/	7975	&	17:40:56.43  & -53:41:46.21	&	2.73		&	22.93	&	23.56	&	0.11	&	21.93	&	0.04	\\
1/	7983	&	17:40:43.36  & -53:41:46.13	&	1.38		&	20.64	&	20.91	&	0.02	&	19.80	&	0.03	\\
1/	7993	&	17:41:03.35  & -53:41:45.59	&	1.45		&	20.35	&	20.48	&	0.03	&	19.43	&	0.04	\\
1/	8023	&	17:40:39.16  & -53:41:45.30 &	1.49		&	19.63	&	20.36	&	0.03	&	19.39	&	0.04	\\
1/	8037	&	17:41:18.20  & -53:41:44.74	&	1.49		&	22.57	&	23.06	&	0.05	&	21.47	&	0.03	\\
1/	8047	&	17:40:49.96  & -53:41:44.77	&	1.90		&	21.79	&	22.25	&	0.05	&	21.20	&	0.07	\\
1/	8102	&	17:40:55.04  & -53:41:43.54	&	2.77		&	22.08	&	22.28	&	0.05	&	20.80	&	0.04	\\
1/	8115	&	17:41:00.55  & -53:41:43.55	&	1.63		&	22.55	&	23.15	&	0.05	&	21.84	&	0.02	\\
1/	8137	&	17:41:15.66  & -53:41:42.25	&	2.05		&	22.21	&	22.42	&	0.04	&	20.90	&	0.06	\\
1/	8153	&	17:41:01.54  & -53:41:42.30	&	1.67		&	22.48	&	23.03	&	0.05	&	21.68	&	0.03	\\
1/	8258	&	17:40:44.95  & -53:41:40.34	&	1.50		&	21.46	&	21.69	&	0.08	&	20.63	&	0.08	\\
1/	8274	&	17:41:01.47  & -53:41:39.16	&	2.24		&	21.82	&	21.86	&	0.08	&	20.57	&	0.07	\\
1/	8285	&	17:41:12.34  & -53:41:39.13	&	1.56		&	21.52	&	21.95	&	0.02	&	20.72	&	0.02	\\
1/	8290	&	17:40:57.16  & -53:41:39.16	&	1.50		&	20.92	&	21.23	&	0.02	&	20.05	&	0.02	\\
1/	8310	&	17:40:49.54  & -53:41:38.50	&	1.78		&	22.30	&	22.98	&	0.08	&	21.52	&	0.05	\\
1/	8350	&	17:40:45.37  & -53:41:37.84	&	2.21		&	21.94	&	22.12	&	0.02	&	21.02	&	0.02	\\
1/	8419	&	17:40:47.00  & -53:41:35.97	&	1.92		&	21.86	&	22.08	&	0.05	&	20.75	&	0.03	\\
1/	8452	&	17:40:51.37  & -53:41:34.75	&	2.19		&	22.32	&	22.98	&	0.07	&	21.36	&	0.04	\\
1/	8465	&	17:41:02.95  & -53:41:33.52	&	1.49		&	22.94	&	23.94	&	0.07	&	22.26	&	0.13	\\
1/	8492	&	17:41:07.05  & -53:41:33.51	&	2.19		&	20.73	&	21.20	&	0.02	&	20.11	&	0.02	\\
1/	8507	&	17:40:53.07  & -53:41:34.13	&	2.75		&	22.33	&	23.16	&	0.11	&	21.56	&	0.03	\\
1/	8529	&	17:40:46.36  & -53:41:33.46	&	1.51		&	18.86	&	19.06	&	0.01	&	18.08	&	0.03	\\
1/	8572	&	17:40:39.16  & -53:41:29.62	&	1.51		&	22.56	&	23.07	&	0.10	&	21.58	&	0.07	\\
1/	8622	&	17:41:01.26  & -53:41:31.01	&	2.19		&	21.97	&	22.67	&	0.05	&	21.29	&	0.04	\\
1/	8639	&	17:40:45.87  & -53:41:30.95	&	2.09		&	21.20	&	21.79	&	0.07	&	20.51	&	0.07	\\
1/	8674	&	17:41:06.55  & -53:41:30.38	&	1.82		&	22.30	&	23.50	&	0.06	&	21.97	&	0.03	\\
1/	8718	&	17:41:15.80  & -53:41:27.83	&	1.36		&	22.80	&	23.62	&	0.06	&	21.96	&	0.02	\\
1/	8732	&	17:40:39.16  & -53:41:32.13	&	1.45		&	20.04	&	20.12	&	0.02	&	19.21	&	0.01	\\
1/	8760	&	17:40:54.83  & -53:41:29.12	&	1.84		&	20.77	&	20.90	&	0.01	&	20.02	&	0.02	\\
1/	8775	&	17:40:50.60  & -53:41:28.47	&	1.91		&	21.82	&	21.97	&	0.04	&	20.85	&	0.02	\\
1/	8817	&	17:40:54.13  & -53:41:27.86	&	2.57		&	22.26	&	22.70	&	0.04	&	21.61	&	0.04	\\
1/	8850	&	17:40:56.74  & -53:41:27.24	&	2.09   	&	19.73	&	20.03	&	0.06	&	19.19	&	0.06	\\
1/	8890	&	17:40:58.79  & -53:41:25.37	&	2.65		&	22.85	&	23.79	&	0.11	&	21.81	&	0.05	\\
1/	8920	&	17:40:53.63  & -53:41:25.98	&	1.75		&	22.68	&	23.62	&	0.12	&	22.38	&	0.18	\\
1/	8926	&	17:40:38.67  & -53:41:25.23	&	2.09		&	21.99	&	22.71	&	0.08	&	21.19	&	0.05	\\
1/	8932	&	17:41:05.64  & -53:41:25.36	&	1.49		&	21.72	&	22.05	&	0.02	&	20.79	&	0.02	\\
1/	8963	&	17:40:54.34  & -53:41:24.48	&	2.28		&	22.15	&	22.69	&	0.04	&	21.42	&	0.03	\\
1/	8982	&	17:41:07.39  & -53:41:23.73	&	1.94		&	21.57	&	21.20	&	0.05	&	20.05	&	0.03	\\
1/	9056	&	17:41:02.22  & -53:41:22.23	&	2.45		&	22.49	&	22.95	&	0.05	&	21.58	&	0.02	\\
1/	9078	&	17:40:51.12  & -53:41:22.20	&	2.49 	&	22.09	&	22.42	&	0.09	&	20.92	&	0.05	\\
1/	9145	&	17:41:24.24  & -53:41:19.84	&	1.42	  	&	21.33	&	22.12	&	0.03	&	20.87	&	0.02	\\
1/	9175	&	17:40:53.66  & -53:41:20.71	&	1.67	 	&	21.68	&	22.07	&	0.02	&	20.75	&	0.01	\\
1/	9187	&	17:40:49.85  & -53:41:19.94	&	2.02		&	22.09	&	21.97	&	0.05	&	20.81	&	0.04	\\
1/	9519	&	17:41:20.12  & -53:45:01.12	&	1.38		&	19.23	&	19.39	&	0.01	&	18.64	&	0.01	\\
1/	9647	&	17:40:38.38  & -53:44:45.26	&	2.24		&	19.00	&	19.13	&	0.01	&	18.28	&	0.03	\\
1/	10089&	17:41:22.15  & -53:44:01.65	&	1.39		&	19.72	&	19.88	&	0.01	&	19.05	&	0.01	\\
1/	10271&	17:40:45.51  & -53:43:41.38	&	1.41		&	19.49	&	19.92	&	0.01	&	19.03	&	0.02	\\
1/	10750&	17:40:52.47  & -53:43:04.55	&	1.32		&	19.28	&	19.54	&	0.02	&	18.66	&	0.02	\\
1/	11485&	17:40:51.46  & -53:42:17.14	&	1.44		&	19.76	&	20.06	&	0.01	&	19.24	&	0.01	\\
\hline
\end{tabular} 
\end{minipage}
\end{table*}

\begin{table*}
\begin{minipage}{180mm}
\caption[]{Continuation. } 
    \label{}
    \vspace{1em}
    \centering
\begin{tabular}{ c c c c c c c  c c c} 
\hline\hline
Star &  R.A. (J2000)    &  Dec. (J2000) &  $\chi^2_{\rm{red}}$  & $\langle$FILT\_465\_250$\rangle$ &  $\langle$B$\_HIGH \rangle$ & $\sigma_B$ &  $\langle$V$\_HIGH \rangle$ &  $\sigma_V$    \\
 ID   & \textit{h m s} &    ${}^{\circ} $ '  "  	  &              &         &		  &       &		         &           \\
\hline 
1/	11550&	17:40:54.17  & -53:42:12.63	&	1.46		&	19.33	&	19.42	&	0.01	&	18.61	&	0.02	\\
1/	11587&	17:41:01.46  & -53:42:10.39	&	1.31		&	19.24	&	19.43	&	0.01	&	18.67	&	0.01	\\
1/	11618&	17:40:51.03  & -53:42:09.61	&	1.86		&	19.56	&	19.74	&	0.02	&	18.90	&	0.03	\\
1/	11759&	17:40:38.66  & -53:42:00.47	&	1.57		&	19.69	&	19.77	&	0.03	&	18.87	&	0.02	\\
1/	12053&	17:40:39.43  & -53:41:45.43	&	1.34		&	17.28	&	16.78	&	0.03	&	16.22	&	0.01	\\
1/	12379&	17:41:03.23  & -53:41:30.51	&	1.38	 	&	19.30	&	19.81	&	0.01	&	18.73	&	0.02	\\ 
1/	12495&	17:40:45.10  & -53:41:25.92	&	1.35 	&	18.39	&	17.53	&	0.09	&	16.95	&	0.11	\\
1/	12590&	17:41:01.71  & -53:41:22.23	&	1.63 	&	19.15	&	19.33	&	0.01	&	18.61	&	0.01	\\
1/	12655&	17:40:46.88  & -53:41:19.92	&	2.57	 	&	19.91	&	19.65	&	0.03	&	18.95	&	0.02	\\
2/	714	&	17:41:08.12  & -53:48:15.51	&	1.90 	&	21.37	&	22.48	&	0.06	&	21.03	&	0.06	\\
2/	911	&	17:41:08.69  & -53:48:08.25 &	1.42 	&	20.00	&	24.21	&	0.06	&	22.52	&	0.02	\\
2/	969	&	17:40:51.82  & -53:48:06.64	&	1.60	 	&	20.03	&	24.15	&	0.07	&	22.50	&	0.03	\\
2/	984	&	17:41:10.17  & -53:48:06.41	&	1.50 	&	19.79	&	24.10	&	0.08	&	22.38	&	0.03	\\
2/	1036	&	17:40:46.53  & -53:48:04.49	&	1.92	 	&	21.46	&	25.49	&	0.21	&	23.91	&	0.06	\\
2/	1040	&	17:41:09.80  & -53:48:04.24	&	1.85	 	&	21.34	&	25.21	&	0.16	&	23.66	&	0.04	\\
2/	1216	&	17:40:43.95  & -53:47:59.78	&	1.46 	&	19.86	&	23.74	&	0.05	&	22.38	&	0.03	\\
2/	1221	&	17:40:58.53  & -53:47:59.31	&	1.75	 	&	21.31	&	25.62	&	0.33	&	23.88	&	0.06	\\
2/	1225	&	17:40:43.46  & -53:47:59.79	&	1.86 	&	21.38	&	25.34	&	0.18	&	23.42	&	0.03	\\
2/	1337	&	17:41:08.16  & -53:47:55.55	&	1.35	 	&	21.07	&	25.98	&	0.32	&	24.26	&	0.09	\\
2/	1339	&	17:40:57.14  & -53:47:55.34	&	1.31	 	&	20.53	&	24.97	&	0.14	&	23.32	&	0.03	\\
2/	1371	&	17:40:50.91  & -53:47:54.67	&	1.31 	&	19.63	&	23.59	&	0.06	&	22.16	&	0.02	\\
2/	1378	&	17:41:09.10  & -53:47:54.09	&	1.31 	&	21.08	&	25.18	&	0.15	&	23.50	&	0.03	\\
2/	1393	&	17:40:43.09  & -53:47:53.98	&	1.44 	&	21.04	&	25.47	&	0.19	&	23.42	&	0.04	\\
2/	1480	&	17:41:00.99  & -53:47:51.30	&	1.36 	&	21.47	&	25.74	&	0.25	&	24.16	&	0.06	\\
2/	1510	&	17:41:04.51  & -53:47:50.53	&	1.45 	&	21.66	&	25.59	&	0.21	&	24.04	&	0.06	\\
2/	1527	&	17:40:45.14  & -53:47:50.34	&	1.43 	&	17.61	&	21.09	&	0.02	&	20.04	&	0.02	\\
2/	1618	&	17:41:01.56  & -53:47:46.94	&	1.67 	&	20.79	&	24.89	&	0.11	&	23.11	&	0.02	\\
2/	1656	&	17:41:16.96  & -53:47:45.96	&	1.38 	&	21.28	&	25.39	&	0.20	&	23.79	&	0.04	\\
2/	1667	&	17:40:40.14  & -53:47:45.64	&	1.84 	&	20.60	&	24.88	&	0.10	&	23.17	&	0.03	\\
2/	1685	&	17:41:06.23  & -53:47:44.34	&	1.83 	&	21.40	&	25.48	&	0.23	&	23.82	&	0.05	\\
2/	1711	&	17:40:54.84  & -53:47:44.11	&	1.41	 	&	20.79	&	24.86	&	0.11	&	23.24	&	0.03	\\
2/	1759	&	17:40:49.68  & -53:47:42.70	&	1.32	 	&	21.69	&	25.58	&	0.21	&	23.93	&	0.04	\\
2/	1872	&	17:40:41.08  & -53:47:39.47	&	1.54 	&	21.29	&	25.73	&	0.26	&	24.13	&	0.06	\\
2/	1958	&	17:41:03.65  & -53:47:35.66	&	1.38	 	&	21.11	&	25.26	&	0.14	&	23.67	&	0.04	\\
2/	1975	&	17:41:23.59  & -53:47:33.11	&	1.34	 	&	19.98	&	24.33	&	0.07	&	22.61	&	0.07	\\
2/	2077	&	17:40:54.84  & -53:47:31.41	&	1.44 	&	19.11	&	22.93	&	0.04	&	21.54	&	0.02	\\
2/	2128	&	17:40:38.42  & -53:47:30.40	&	1.83 	&	20.23	&	24.31	&	0.09	&	22.76	&	0.04	\\
2/	2152	&	17:41:02.66  & -53:47:29.14	&	1.46	 	&	20.85	&	24.95	&	0.18	&	23.13	&	0.03	\\
2/	2181	&	17:40:42.84  & -53:47:30.03	&	1.74	 	&	21.52	&	25.78	&	0.30	&	23.72	&	0.04	\\
2/	2214	&	17:41:07.01  & -53:47:27.27	&	1.86 	&	21.17	&	25.78	&	0.24	&	24.91	&	0.13	\\
2/	2226	&	17:40:54.72  & -53:47:27.78	&	1.37 	&	20.53	&	24.77	&	0.13	&	23.05	&	0.03	\\
2/	2312	&	17:40:42.35  & -53:47:24.59	&	1.41 	&	20.16	&	23.93	&	0.05	&	22.44	&	0.02	\\
2/	2408	&	17:40:50.21  & -53:47:22.01	&	1.59 	&	21.44	&	25.69	&	0.23	&	23.83	&	0.04	\\
2/	2426	&	17:41:09.09  & -53:47:21.43	&	1.59		&	20.20	&	24.34	&	0.08	&	22.71	&	0.06	\\
2/	2469	&	17:41:11.96  & -53:47:19.56	&	1.37		&	19.61	&	23.66	&	0.04	&	22.11	&	0.02	\\
2/	2478	&	17:40:40.79  & -53:47:19.51	&	1.30		&	21.03	&	25.58	&	0.21	&	23.84	&	0.07	\\
2/	2483	&	17:40:47.92  & -53:47:19.85	&	1.38		&	21.11	&	25.33	&	0.21	&	23.55	&	0.05	\\
2/	2516	&	17:41:12.25  & -53:47:18.47	&	1.88		&	19.64	&	23.30	&	0.07	&	21.99	&	0.03	\\
2/	2567	&	17:40:49.03  & -53:47:17.30	&	1.35		&	20.88	&	25.13	&	0.14	&	23.62	&	0.04	\\
2/	2580	&	17:40:52.88  & -53:47:16.91	&	1.63		&	21.45	&	25.81	&	0.25	&	23.79	&	0.05	\\
2/	2610	&	17:40:39.81  & -53:47:15.88	&	1.52		&	20.68	&	24.75	&	0.09	&	22.95	&	0.03	\\
2/	2661	&	17:40:48.41  & -53:47:14.40	&	1.84		&	20.89	&	24.56	&	0.09	&	23.14	&	0.04	\\
2/	2677	&	17:40:47.06  & -53:47:14.05	&	2.04		&	20.75	&	24.65	&	0.11	&	23.19	&	0.04	\\
2/	2746	&	17:40:55.00  & -53:47:12.17	&	2.46		&	20.85	&	24.90	&	0.23	&	23.23	&	0.06	\\
2/	2939	&	17:40:55.49  & -53:47:06.36	&	1.59		&	20.33	&	24.44	&	0.08	&	22.77	&	0.03	\\
2/	2955	&	17:40:44.23  & -53:47:02.45	&	2.01		&	21.13	&	24.90	&	0.12	&	23.29	&	0.04	\\
2/	2980	&	17:41:18.14  & -53:47:04.93	&	1.82		&	20.94	&	25.15	&	0.14	&	23.43	&	0.04	\\
2/	3031	&	17:41:21.17  & -53:47:02.68	&	1.33		&	20.23	&	24.73	&	0.10	&	23.02	&	0.03	\\
2/	3060	&	17:40:44.19  & -53:47:06.08	&	1.96		&	20.40	&	24.07	&	0.07	&	22.74	&	0.03	\\
2/	3098	&	17:41:18.55  & -53:47:00.56	&	1.40		&	20.41	&	24.54	&	0.08	&	22.90	&	0.03	\\
2/	3135	&	17:41:14.25  & -53:46:59.92	&	1.39		&	21.59	&	25.85	&	0.28	&	23.99	&	0.05	\\
\hline
\end{tabular} 
\end{minipage}
\end{table*}

\begin{table*}
\begin{minipage}{180mm}
\caption[]{Continuation. } 
    \label{}
    \vspace{1em}
    \centering
\begin{tabular}{ c c c c c c c  c c c} 
\hline\hline
Star &  R.A. (J2000)    &  Dec. (J2000) &  $\chi^2_{\rm{red}}$  & $\langle$FILT\_465\_250$\rangle$ &  $\langle$B$\_HIGH \rangle$ & $\sigma_B$ &  $\langle$V$\_HIGH \rangle$ &  $\sigma_V$    \\
 ID   & \textit{h m s} &    ${}^{\circ} $ '  "  	  &              &         &		  &       &		         &           \\
\hline
2/	3156	&	17:41:19.69  & -53:46:58.72	&	1.67		&	21.05	&	25.03	&	0.13	&	23.55	&	0.04	\\
2/	3171	&	17:40:41.45  & -53:46:58.83	&	1.97		&	20.49	&	24.53	&	0.08	&	22.85	&	0.03	\\
2/	3213	&	17:40:38.71  & -53:46:58.83	&	2.07		&	21.06	&	25.07	&	0.15	&	23.37	&	0.04	\\
2/	3214	&	17:41:20.59  & -53:46:57.25	&	1.54		&	20.76	&	24.93	&	0.12	&	23.08	&	0.03	\\
2/	3314	&	17:40:40.63  & -53:46:54.84	&	1.58		&	21.51	&	25.66	&	0.25	&	23.97	&	0.05	\\
2/	3340	&	17:41:10.11  & -53:46:53.83	&	1.54		&	20.17	&	24.14	&	0.06	&	22.54	&	0.02	\\
2/	3344	&	17:41:21.58  & -53:46:53.60	&	1.38		&	19.56	&	23.24	&	0.07	&	21.78	&	0.04	\\
2/	3352	&	17:40:43.99  & -53:46:53.74	&	1.62		&	18.91	&	22.72	&	0.03	&	21.41	&	0.02	\\
2/	3443	&	17:40:41.53  & -53:46:51.20	&	1.38		&	20.53	&	24.62	&	0.09	&	23.01	&	0.02	\\
2/	3466	&	17:40:48.94  & -53:46:50.08	&	1.53		&	21.41	&	25.35	&	0.18	&	23.64	&	0.04	\\
2/	3498	&	17:41:11.83  & -53:46:49.81	&	1.32		&	19.00	&	22.59	&	0.03	&	21.61	&	0.02	\\
2/	3534	&	17:41:10.44  & -53:46:48.02	&	1.84		&	20.44	&	24.70	&	0.13	&	22.87	&	0.03	\\
2/	3575	&	17:41:20.75  & -53:46:45.63	&	1.68		&	20.66	&	25.49	&	0.20	&	23.88	&	0.05	\\
2/	3605	&	17:40:47.47  & -53:46:46.46	&	1.69		&	20.41	&	24.24	&	0.06	&	22.67	&	0.02	\\
2/	3608	&	17:41:14.37  & -53:46:46.13	&	1.39		&	21.23	&	25.26	&	0.13	&	23.74	&	0.04	\\
2/	3658	&	17:40:53.53  & -53:46:44.61	&	1.88		&	21.08	&	24.98	&	0.13	&	23.34	&	0.03	\\
2/	3673	&	17:41:04.79  & -53:46:44.12	&	2.21		&	20.27	&	24.33	&	0.07	&	22.61	&	0.02	\\
2/	3676	&	17:40:49.47  & -53:46:44.28	&	1.46		&	20.51	&	25.33	&	0.28	&	23.40	&	0.08	\\
2/	3678	&	17:40:45.63  & -53:46:43.93	&	1.57		&	19.27	&	23.18	&	0.04	&	21.53	&	0.02	\\
2/	3709	&	17:41:19.98  & -53:46:42.02	&	1.45		&	20.93	&	24.88	&	0.12	&	23.33	&	0.04	\\
2/	3733	&	17:40:38.46  & -53:46:42.50	&	1.91		&	21.01	&	25.07	&	0.15	&	23.23	&	0.04	\\
2/	3815	&	17:41:00.20  & -53:46:38.37	&	1.63		&	20.65	&	25.13	&	0.15	&	23.58	&	0.06	\\
2/	3864	&	17:41:05.24  & -53:46:37.94	&	1.45		&	20.41	&	24.81	&	0.10	&	22.91	&	0.02	\\
2/	3866	&	17:40:41.45  & -53:46:38.14	&	1.64		&	19.55	&	23.49	&	0.05	&	22.03	&	0.02	\\
2/	4018	&	17:41:05.07  & -53:46:32.86	&	1.45		&	21.27	&	25.78	&	0.25	&	23.83	&	0.04	\\
2/	4054	&	17:40:48.45  & -53:46:31.58	&	1.55		&	21.54	&	25.69	&	0.26	&	23.82	&	0.05	\\
2/	4056	&	17:41:24.80  & -53:46:31.02	&	2.05		&	19.00	&	23.54	&	0.10	&	22.34	&	0.05	\\
2/	4059	&	17:41:12.85  & -53:46:31.28	&	1.38		&	20.73	&	24.91	&	0.10	&	23.28	&	0.04	\\
2/	4063	&	17:40:53.98  & -53:46:31.54	&	1.32		&	21.08	&	25.05	&	0.14	&	23.49	&	0.04	\\
2/	4109	&	17:41:02.98  & -53:46:29.63	&	1.64		&	19.58	&	23.39	&	0.06	&	21.94	&	0.03	\\
2/	4164	&	17:40:50.21  & -53:46:28.30	&	1.60		&	22.00	&	25.44	&	0.20	&	23.92	&	0.05	\\
2/	4167	&	17:41:23.86  & -53:46:27.78	&	1.68		&	20.98	&	24.92	&	0.14	&	23.21	&	0.03	\\
2/	4205	&	17:40:49.27  & -53:46:26.86	&	1.45		&	19.82	&	23.94	&	0.06	&	22.32	&	0.03	\\
2/	4219	&	17:41:09.78  & -53:46:26.62	&	1.42		&	21.70	&	25.57	&	0.22	&	23.70	&	0.04	\\
2/	4292	&	17:41:22.10  & -53:46:24.19	&	1.61		&	21.54	&	25.71	&	0.26	&	23.87	&	0.05	\\
2/	4321	&	17:41:09.04  & -53:46:24.09	&	1.31		&	21.00	&	25.45	&	0.20	&	23.47	&	0.03	\\
2/	4356	&	17:41:24.06  & -53:46:22.69	&	1.47		&	19.85	&	23.91	&	0.05	&	22.32	&	0.02	\\
2/	4372	&	17:40:50.41  & -53:46:22.50	&	1.93		&	21.66	&	25.23	&	0.13	&	23.67	&	0.04	\\
2/	4374	&	17:40:46.20  & -53:46:22.52	&	1.65		&	20.93	&	25.18	&	0.15	&	23.52	&	0.03	\\
2/	4379	&	17:40:59.75  & -53:46:22.41	&	1.81		&	21.51	&	25.05	&	0.14	&	23.42	&	0.04	\\
2/	4390	&	17:40:51.89  & -53:46:22.12	&	1.34		&	20.26	&	24.44	&	0.07	&	22.88	&	0.02	\\
2/	4393	&	17:40:40.22  & -53:46:22.54	&	1.74		&	20.51	&	24.59	&	0.12	&	22.73	&	0.02	\\
2/	4413	&	17:40:48.98  & -53:46:22.14	&	1.33		&	21.26	&	25.51	&	0.20	&	23.79	&	0.07	\\
2/	4426	&	17:41:11.90  & -53:46:20.78	&	1.36		&	21.33	&	25.56	&	0.21	&	23.80	&	0.05	\\
2/	4508	&	17:40:44.07  & -53:46:19.26	&	1.68		&	20.95	&	25.31	&	0.23	&	23.43	&	0.05	\\
2/	4617	&	17:40:49.35  & -53:46:27.58	&	1.61		&	20.60	&	24.85	&	0.12	&	23.25	&	0.03	\\
2/	4663	&	17:41:06.34  & -53:46:24.86	&	1.41		&	20.74	&	25.02	&	0.14	&	23.35	&	0.04	\\
2/	4706	&	17:40:42.39  & -53:46:13.46	&	1.74		&	21.67	&	25.48	&	0.20	&	23.58	&	0.04	\\
2/	4715	&	17:40:51.68  & -53:46:13.05	&	1.62		&	20.95	&	24.62	&	0.08	&	23.13	&	0.03	\\
2/	4798	&	17:41:02.24  & -53:46:10.40	&	1.33		&	20.43	&	24.76	&	0.11	&	23.05	&	0.03	\\
2/	4799	&	17:40:56.43  & -53:46:10.47	&	1.58		&	20.11	&	24.46	&	0.08	&	22.98	&	0.04	\\
2/	4839	&	17:41:13.58  & -53:46:09.13	&	1.33		&	20.75	&	25.16	&	0.14	&	23.17	&	0.03	\\
2/	4907	&	17:41:07.77  & -53:46:07.78	&	2.27		&	19.66	&	22.15	&	0.10	&	20.65	&	0.14	\\
2/	4919	&	17:40:56.76  & -53:46:06.84	&	1.40		&	21.18	&	25.10	&	0.13	&	23.52	&	0.05	\\
2/	4920	&	17:40:52.13  & -53:46:06.88	&	1.32		&	20.05	&	24.05	&	0.17	&	22.76	&	0.11	\\
2/	5028	&	17:41:01.34  & -53:46:03.88 &	1.41		&	20.18	&	24.54	&	0.10	&	22.65	&	0.02	\\
2/	5080	&	17:41:06.78  & -53:46:02.35	&	1.50		&	20.55	&	24.32	&	0.10	&	22.75	&	0.03	\\
2/	5107	&	17:41:23.32  & -53:46:01.30	&	1.31		&	19.87	&	23.77	&	0.05	&	22.36	&	0.03	\\
2/	5126	&	17:40:41.37  & -53:46:01.13	&	1.41		&	20.88	&	24.66	&	0.10	&	23.23	&	0.03	\\
2/	5176	&	17:41:16.77  & -53:45:59.27 &	1.71		&	19.52	&	23.77	&	0.30	&	22.00	&	0.06	\\
2/	5211	&	17:40:51.43  & -53:45:58.54	&	1.50		&	21.50	&	25.44	&	0.32	&	23.68	&	0.06	\\
2/	5232	&	17:40:57.08  & -53:45:57.76	&	1.43		&	19.98	&	24.23	&	0.09	&	22.72	&	0.04	\\
\hline
\end{tabular} 
\end{minipage}
\end{table*}

\begin{table*}
\begin{minipage}{180mm}
\caption[]{Continuation. } 
    \label{}
    \vspace{1em}
    \centering
\begin{tabular}{ c c c c c c c  c c c} 
\hline\hline
Star &  R.A. (J2000)    &  Dec. (J2000) &  $\chi^2_{\rm{red}}$  & $\langle$FILT\_465\_250$\rangle$ &  $\langle$B$\_HIGH \rangle$ & $\sigma_B$ &  $\langle$V$\_HIGH \rangle$ &  $\sigma_V$    \\
 ID   & \textit{h m s} &    ${}^{\circ} $ '  "  	  &              &         &		  &       &		         &           \\
\hline 
2/	5243	&	17:40:50.82  & -53:45:57.09	&	1.69		&	19.18	&	23.11	&	0.06	&	21.75	&	0.02	\\
2/	5274	&	17:41:12.39  & -53:45:54.64	&	1.52		&	21.09	&	25.90	&	0.27	&	23.88	&	0.06	\\
2/	5364	&	17:41:15.95  & -53:45:53.12	&	1.33		&	21.36	&	25.34	&	0.17	&	23.66	&	0.04	\\
2/	5384	&	17:41:09.52  & -53:45:53.24	&	1.43		&	19.58	&	23.62	&	0.07	&	22.21	&	0.04	\\
2/	5440	&	17:40:43.29  & -53:45:52.05	&	1.60		&	19.34	&	23.41	&	0.07	&	22.12	&	0.05	\\
2/	5500	&	17:41:13.62  & -53:45:49.90	&	1.98		&	21.47	&	25.31	&	0.17	&	23.39	&	0.06	\\
2/	5540	&	17:40:56.18  & -53:45:48.70	&	1.38		&	20.78	&	24.80	&	0.10	&	23.07	&	0.03	\\
2/	5552	&	17:40:42.47  & -53:45:48.06	&	1.65		&	20.44	&	24.59	&	0.09	&	22.92	&	0.03	\\
2/	5553	&	17:41:24.71  & -53:45:47.48	&	1.41		&	18.91	&	22.87	&	0.04	&	21.46	&	0.02	\\
2/	5678	&	17:40:55.57  & -53:45:45.07	&	1.60		&	21.00	&	25.28	&	0.18	&	23.64	&	0.04	\\
2/	5705	&	17:41:13.98  & -53:45:43.36	&	1.37		&	18.38	&	22.14	&	0.07	&	20.62	&	0.08	\\
2/	5718	&	17:40:49.76  & -53:45:43.31	&	1.54		&	21.15	&	25.22	&	0.15	&	23.58	&	0.04	\\
2/	5730	&	17:40:59.74  & -53:45:42.85	&	1.52		&	20.93	&	25.20	&	0.18	&	23.35	&	0.04	\\
2/	5780	&	17:41:12.80  & -53:45:41.20	&	1.74		&	20.67	&	24.59	&	0.11	&	23.11	&	0.04	\\
2/	5827	&	17:41:23.56  & -53:45:39.88	&	1.35		&	19.00	&	22.89	&	0.04	&	21.53	&	0.01	\\
2/	5840	&	17:40:39.16  & -53:45:40.08	&	1.51		&	17.49	&	21.20	&	0.02	&	20.17	&	0.02	\\
2/	5855	&	17:40:55.12  & -53:45:39.27	&	1.87		&	20.00	&	23.95	&	0.06	&	22.32	&	0.02	\\
2/	5884	&	17:41:04.12  & -53:45:38.44	&	1.63		&	20.05	&	24.03	&	0.05	&	22.58	&	0.04	\\
2/	5919	&	17:40:46.32  & -53:45:37.88	&	1.79		&	20.28	&	24.51	&	0.11	&	22.72	&	0.02	\\
2/	5937	&	17:41:02.36  & -53:45:37.01	&	1.33		&	21.34	&	25.87	&	0.31	&	23.72	&	0.04	\\
2/	5957	&	17:40:51.43  & -53:45:36.40	&	2.21		&	21.30	&	25.51	&	0.23	&	23.59	&	0.08	\\
2/	6006	&	17:40:55.77  & -53:45:34.91	&	1.92		&	20.15	&	24.58	&	0.10	&	22.75	&	0.03	\\
2/	6007	&	17:40:51.88  & -53:45:35.67	&	2.27		&	20.86	&	25.39	&	0.25	&	23.82	&	0.11	\\
2/	6014	&	17:40:49.88  & -53:45:34.60	&	1.96		&	21.76	&	25.23	&	0.19	&	23.73	&	0.06	\\
2/	6036	&	17:40:54.46  & -53:45:34.56	&	1.44		&	19.77	&	23.91	&	0.07	&	22.26	&	0.02	\\
2/	6151	&	17:41:21.47  & -53:45:30.86	&	1.81		&	20.26	&	24.45	&	0.09	&	22.66	&	0.02	\\
2/	6189	&	17:40:43.37  & -53:45:30.27	&	1.62		&	19.73	&	23.95	&	0.05	&	22.24	&	0.02	\\
2/	6229	&	17:40:55.69  & -53:45:29.11	&	1.49		&	19.67	&	22.59	&	0.10	&	20.96	&	0.09	\\
2/	6230	&	17:40:50.57  & -53:45:29.51	&	1.41		&	20.27	&	24.36	&	0.07	&	22.65	&	0.02	\\
2/	6248	&	17:41:07.60  & -53:45:28.59	&	1.50		&	20.06	&	24.26	&	0.09	&	22.52	&	0.06	\\
2/	6260	&	17:40:57.90  & -53:45:28.36	&	1.39		&	20.76	&	25.01	&	0.16	&	23.24	&	0.03	\\
2/	6291	&	17:41:17.66  & -53:45:27.32	&	1.84		&	21.15	&	25.44	&	0.24	&	23.68	&	0.09	\\
2/	6300	&	17:40:45.70  & -53:45:27.36	&	1.50		&	21.12	&	19.50	&	0.01	&	18.71	&	0.00	\\
2/	6327	&	17:41:16.93  & -53:45:25.88	&	2.17		&	21.70	&	25.51	&	0.24	&	23.78	&	0.06	\\
2/	6352	&	17:41:00.93  & -53:45:25.78	&	1.80		&	20.80	&	25.35	&	0.23	&	23.37	&	0.09	\\
2/	6357	&	17:40:53.93  & -53:45:25.13	&	2.18		&	20.19	&	24.79	&	0.20	&	22.82	&	0.03	\\
2/	6376	&	17:41:07.02  & -53:45:24.25	&	1.31		&	20.23	&	24.68	&	0.17	&	22.78	&	0.04	\\
2/	6386	&	17:41:22.98  & -53:45:23.93	&	1.70		&	20.07	&	24.17	&	0.07	&	22.43	&	0.02	\\
\hline
\end{tabular} 
\end{minipage}
\end{table*}

\end{document}